\documentclass[aps,prl,twocolumn,showpacs,groupedaddress,preprintnumbers]{revtex4-2}  

\usepackage{graphicx}  
\usepackage{dcolumn}   
\usepackage{bm}        
\usepackage{amssymb}   
\usepackage{braket}
\usepackage{amsmath}
\usepackage{ulem}
\usepackage{color}  	

\usepackage{bbm}
\usepackage{slashed}
\usepackage{subfigure}
\usepackage[utf8]{inputenc}
\DeclareGraphicsExtensions{.pdf,.png,.jpg,.eps}
\graphicspath{ {./images/} }

\newcommand{\Tr}{\mathrm{Tr}}

\newcommand{\td}{\mathrm{d}}

\newcommand{\TT}[1]{\mathrm{#1}}

\renewcommand{\v}[1]{\mathbf{#1}}

\newcommand{\Nc}{N_{\mathrm{c}}}
\newcommand{\As}{\alpha_{\mathrm{s}}}

\begin{document}

\title{The one-jettiness distribution contains super-super-leading logarithms}

\author{Andrea Banfi}
\email{A.Banfi@sussex.ac.uk}
\affiliation{University of Sussex, Department of Physics and Astronomy, Brighton BN1 9RH, UK }

\author{Jeffrey R. Forshaw}
\email{jeffrey.forshaw@manchester.ac.uk}
\affiliation{Department of Physics and Astronomy, University of Manchester, Manchester M13 9PL, United Kingdom}

\author{Jack Holguin}
\email{jack.holguin@manchester.ac.uk}
\affiliation{Department of Physics and Astronomy, University of Manchester, Manchester M13 9PL, United Kingdom}

\date{\today}

\begin{abstract}
We show that one-jettiness ($\tau_1$) in colour-singlet plus jet production suffers from super-leading logarithms starting at order $\As^4 \ln(1/\tau_1)^6$ relative to the Born level. This is one logarithm more dominant than any previously identified super-leading logarithms. The extra logarithm is not associated with additional poles, and is therefore consistent with the factorization of universal parton distribution functions at scale $\tau_1 Q$, where $Q$ is the hard scale.
\end{abstract}


\maketitle
\section{Introduction}
$N$-jettiness \cite{Stewart:2010tn} is a widely studied observable that is sensitive to QCD radiation. It is also used as a tool to handle soft singularities in NNLO calculations \cite{Gaunt:2015pea,Boughezal:2015dva,Boughezal:2015aha,Boughezal:2016wmq,Moult:2017jsg,Mondini:2019gid} owing to its calculability at fixed order and apparently simple logarithmic structure. According to \cite{Alioli:2023rxx}, one-jettiness in colour-singlet (e.g. Higgs or $Z$) plus jet production at hadron colliders is free from both non-global 
\cite{non_global_logs} and super-leading logarithms \cite{Forshaw:2006fk}, and this facilitates a resummed calculation at N$^3$LL accuracy. In this letter we show that this is not the case and that, in fact, one-jettiness ($\tau_1$) is an interesting source of super-leading logarithms. Extraordinarily, these logarithms start at order $\As^4 L^6$ relative to the lowest order colour-singlet plus jet process, where $L = \ln( 1/\tau_1)$. This is a next-to-next-to double-log effect and so could have a direct impact on LHC studies that utilise this observable, including dark matter searches \cite{Lindert:2017olm}, and Higgs and electroweak physics \cite{Jouttenus:2013hs}.

We take the one-jettiness of a hadron-hadron collision event to be
\begin{align}
\tau_1 = \frac{1}{Q}\sum_k \min \left\{ 
n_a \cdot p_k, \,
n_b \cdot p_k, \,
n_J \cdot p_k
\right\},
\end{align}
where $n_a$, $n_b$ and $n_J$ are light-like vectors along the directions of the colliding beams and the jet. $Q$ is the hard scale, e.g. the transverse momentum of the $H/Z$. The index $k$ sums over all final state particles. Taking $\tau_1 \to 0$ is the limit that all radiation lies within the beam remnants or the jet, i.e. there is a veto on out-of-jet radiation. For simplicity, we will focus our attention on the process $qg \to q Z$. We label the partons $ab \to c Z$ where $a$ may be either the quark or the gluon.

With a fixed QCD coupling ($\As$), the naive expectation is that the cumulative distribution of the one-jettiness obeys the form,
\begin{align}
   \sigma(\tau_1) = \int_0^{\tau_1} \frac{\td \sigma}{\td \tau_1'} \td \tau_1' &= \sigma \exp\left( -\frac{\As}{2\pi} L^2 (2 C_F + C_A)\right)  \nonumber \\ & \hspace{-3 cm} \times\Big[ g_0(\As L) + \As g_1(\As L) 
+ \As^2 g_2(\As L) + \cdots \Big]. \label{eq:logs}
\end{align} 
As we will see, since the super-leading logarithms start at $\alpha_s^4 L^6$, even the complete exponentiation of the double logarithms is spoiled and the observable cannot be written in the form of Eq.~\eqref{eq:logs}. \vspace*{0.2cm}

\section{Super-leading logarithms}

Super-leading logarithms result from a violation of QCD coherence due to the non-trivial colour associated with Coulomb gluon exchange, and they amount to a failure of the DGLAP `plus prescription'. As discussed in \cite{Banfi:2010xy,Forshaw:2021fxs}, the effect is not always super-leading and so the relevant logarithms are known more generally as coherence violating logarithms (CVL). The lowest order coherence violating contribution to one-jettiness in $A+B \rightarrow {\rm jet} + Z$ arises from a soft-collinear emission off one of the incoming legs and, if the emission is off parton $a$, can be written \cite{Forshaw:2021fxs}  
\begin{align}
    \frac{\td \sigma^{\rm CVL}_1}{\td x_a \td x_b \, \td \mathcal{B}} = &\sum_{a = q,g} \frac{\As}{\pi} f^a_A(x_a,\mu_{\TT{F}}) f^b_B(x_b,\mu_{\rm F})   ~    \nonumber \\
    & \times \int_{\mu_{\TT{F}}}^Q \frac{\td k_T}{k_T} \int_{x_a}^{1-\frac{k_T}{Q}} \td z \, \frac{2}{1-z} \; u(k) \,  \nonumber \\ 
    &  \times \,  W_{k_T,Q} \; \text{Tr} (\v{V}_{\mu_{\TT{F}},k_T}^\dag \v{V}_{\mu_{\TT{F}},k_T} \v{t}_{a} \v{H}_{ab}(\mathcal{B})\v{t}^{\dagger}_{a}), \label{eq:master}
\end{align}
where $a,b$ are incoming partons from hadrons $A,B$ and $f^{a/b}_{A/B}$ are the parton distribution functions. $\mu_\text{F}$ is a factorisation scale and $W_{k_T,Q}$ is the Sudakov factor for non-emission off the primary partons, i.e. the exponential pre-factor in Eq.~\eqref{eq:logs} with $L \to \ln (k_T / \tau_1 Q$). $\v{H}_{ab}$ is the Born-level hard density-matrix with phase-space $\mathcal{B}$, the factor $2/(1-z)$ is the soft limit of the $P_{a\rightarrow a}(z)$ collinear splitting function.  $\v{t}_{a}$ is the colour charge operator which acts on the 3 parton colour state to make a 4 parton 

state containing gluon $d$ emitted from parton $a$ \cite{SuperleadingLogs}. Its quadratic form returns the Casimir, i.e. $\v{t}_{a}^{\dagger}\v{t}_{a}=C_{a} \v{1}_3$. We work throughout with fixed $\As$, which is sufficient to compute the leading CVL. The Sudakov operator is
\pagebreak 
\begin{widetext}
\begin{align} 
	&\v{V}_{\mu_{\TT{F}},k_T} \approx {\rm Pexp} \left( \frac{\As}{\pi}  \left[ \sum_{i < j} \v{T}_{i} \cdot \v{T}_{j} \int_{\mu_{\TT{F}}}^{k_T} \frac{\td  q^{(ij)}_T}{q^{(ij)}_T}  \int_{-\ln Q/q^{(ij)}_T}^{\ln Q/q^{(ij)}_T} \td y^{(ij)} \int_0^{2\pi}\frac{\td \phi^{(ij)}}{2\pi}~ \big(1 - u_2(q,k)\big) - i\pi \v{T}_s^2 \ln \frac{k_T}{\mu_{\TT{F}}} \right] \right), \label{eq:V}
\end{align}
\end{widetext}
where $\mathbf{T}_s^2 = (\mathbf{T}_a + \mathbf{T}_b)^2$. Here, colour charges $\v{T}_{i}$ act on parton $i\in\{a,b,c,d\}$ in the 4 parton system. $(q_T^{(ij)}, y^{(ij)},\phi^{(ij)})$ are the transverse momentum, rapidity and azimuth defined in the zero momentum frame of partons $i$ and $j$. 

The one-jettiness is implemented via the measurement function: 
\begin{align}
    u(k) &= \Theta\left(\tau_1 - \frac{1}{Q}\min(n_a \cdot k,n_b \cdot k,n_c \cdot k) \right) \nonumber \\
    u(k_1,k_2) &= \Theta\left(\tau_1 - \sum_{i=1,2}\frac{1}{Q}\min(n_a \cdot k_i,n_b \cdot k_i,n_c \cdot k_i) \right),
\end{align}
where $n_c$ is the lightlike vector along the direction of the outgoing quark, and $u_2(q,k)$ is defined such that $u(k,q) = u_2(q,k) u(k)$.
For the leading CVL, we take the collinear limit (recalling that $k$ is collinear to incoming parton $a$) and make the multiple uncorrelated emission approximation:
\begin{align}
    &u(k) \approx \Theta\left(\tau_1 - \frac{k_T}{Q}e^{-y_k} \right) \nonumber \\
    &u_2(q,k) \approx \Theta\left(\tau_1 - \frac{1}{Q}\min(n_a \cdot q,n_b \cdot q,n_c \cdot q) \right),
\end{align}
where $y_k = - \ln \tan (\theta /2)$ is the rapidity of $k$ in the lab frame (the $a$ direction corresponding to $y_k \to \infty$). These measurement functions are inclusive of all radiation below $k_t \lesssim \tau_1 Q$ and so we set $\mu_{\rm F} = \tau_1 Q$.

\begin{figure*}[t]
  \centering
  \includegraphics[width=0.35\textwidth]{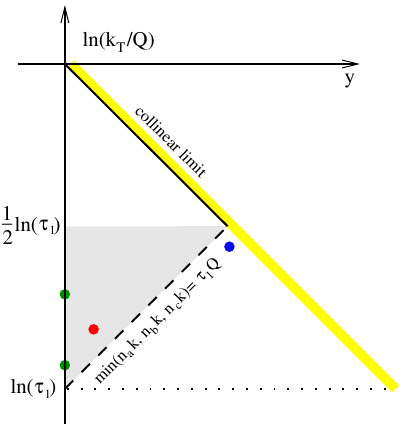} ~~~~~~~~~~~~~
  \raisebox{1cm}{\includegraphics[width=0.45\textwidth]{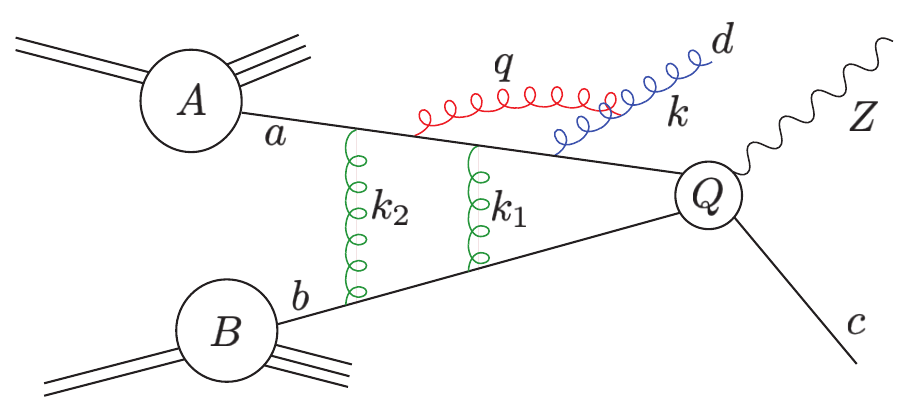}}
  \caption{Left: the Lund plane indicating the region over which the soft gluon (red) is integrated (shaded grey). The collinear gluon (blue) is integrated over the lower (unshaded) triangle. The green dots dots indicate the Coulomb gluons. Right: the corresponding diagram with the momenta labelled as in the text. Here we consider one of the possible contributions corresponding to the indicated ordering of transverse momenta.}
  \label{fig:LundPlane}
\end{figure*}

Eq.~\eqref{eq:master} can be simplified by performing the integrals inside $\v{V}_{\mu_{\rm F}, k_{T}}$. To do this, we will assume that the outgoing quark is well separated from the beam directions. The soft gluon integral is over the domain where real gluon emission would be forbidden, i.e. the measurement function enforces an integral over a sphere with three holes, each centred on the $n_a$, $n_b$ and $n_c$ directions. For dipoles $(a,c),(b,d),(b,c)$ the dipole frame is equivalent to the lab frame, modulo only moderate boosts, and the holes have angular radius $2\tau_1 Q / q_T^{(ij)}.$  Also, for each dipole we may neglect the hole that is not centred on the dipole axis, e.g. for the $(b,d)$ dipole we can neglect the hole centred on parton $c$. The rapidity integral then gives a factor $2\ln(q_T^{(ij)}/(\tau_1 Q))$. This is sufficient for the leading logarithmic result though we can also perform the integral without approximation and obtain (see Eq.~(4.21) of  \cite{Banfi:2003jj}):
\begin{align}
    \int_{q.n_{i,j} > \tau_1 Q} \td y^{(ij)}  \frac{\td \phi^{(ij)}}{2\pi } &= \ln \left(\frac{n_i \cdot n_j}{1 - \frac{q_T^2-\tau_1^2 Q^2}{q_T^2+\tau_1^2 Q^2}} -1 \right) \nonumber \\
    &\approx \ln \left( \frac{q_T^2}{\tau_1^2 Q^2} \frac{n_i \cdot n_j}{2}\right)~,
\end{align}
where $q_T \equiv q_T^{(ij)}$.
Let us make an observation before moving on to consider the $(a,d)$ dipole. We have found an additional logarithm of $q_T/\tau$ and the typical soft wide-angle logarithm of $n_i \cdot n_j$. We might expect the additional logarithm to factor from the $i\pi$ terms by becoming diagonal through colour conservation. In which case, it would not result in a SLL. However, noting the $\v{T}_i \cdot \v{T}_j$ colour factors in Eq.~\eqref{eq:V}, this will only occur if dipole $(a,d)$ also contains the same logarithm.

Dipole $(a,d)$ has an opening angle which scales with $\sqrt{\tau}$ and so the dipole frame differs from the lab frame by a logarithmically large boost. In the dipole frame, the observable is distorted so that almost the entire sphere is within the hole centred on $n_a$ and soft radiation from the $(a,d)$ dipole is only allowed into a small backwards region of size $~\sqrt{\tau}$. Therefore, we expect a contribution from the $(a,d)$ dipole that is power suppressed. To compute this, we can again ignore the hole in the direction of out-going quark $c$ whilst the other two legs are within the hole centred on $n_a$. We now find that (see Eq.~(C.2) in \cite{Banfi:2002hw})
\begin{align}
    \int_{q.n_{a,b} > \tau_1 Q} \! \! \td y^{(ad)}  \frac{\td \phi^{(ad)}}{2\pi } &\! = \! \frac{1}{2}\ln \left(\frac{\tau_1^2 Q^2}{q_T^2} \frac{\cos{\theta}+\frac{q_T^2-\tau_1^2 Q^2}{q_T^2+\tau_1^2 Q^2}}{\cos{\theta}-\frac{q_T^2-\tau_1^2 Q^2}{q_T^2+\tau_1^2 Q^2}}\right) \! .
\end{align}
Importantly, this vanishes as $\theta \rightarrow 0$ and is indeed power suppressed.

Keeping only the leading-logarithmic results, we find that
\begin{align} 
	&\v{V}_{\mu_{\TT{F}},k_T} \approx \nonumber \\
    &  {\rm exp}\!\! \left( \! \frac{\As}{\pi} \! \!  \left[ \sum_{\substack{i < j \\ i \nparallel j } } \v{T}_{i} \cdot \v{T}_{j} \!\!\int_{\mu_{\TT{F}}}^{k_T} \!\frac{\td  q_T}{q_T} 2 \ln\left(\!\frac{q_T}{\tau_1 Q}\!\right)\!\! -\! i\pi \v{T}_s^2 \ln \frac{k_T}{\mu_{\TT{F}}} \right] \!\right),
\end{align}
where the matrix structure of exponential has factorised from the $q_T$ integral and so the path ordering can be dropped. The sum is now restricted to $i < j$ and also parton $i$ not being collinear to parton $j$ ($i \nparallel j$). Since we have dropped the path ordering, we can further simplify the Sudakov:
\begin{align} 
	&\v{V}_{\mu_{\TT{F}},k_T} \approx {\rm exp} \left( \frac{\As}{\pi}  \left[ \v{T}_{ad}^2 L_2(\mu_{\rm F} , k_T) - i\pi \v{T}_s^2 \ln \frac{k_T}{\mu_{\TT{F}}} \right] \right),
\end{align}
where
\begin{align}
    \v{T}_{ad}^2 &= \sum_{\substack{i < j \\ i \nparallel j } } \v{T}_{i} \cdot \v{T}_{j} = - \frac{1}{2} \sum_{i} \v{T}_{i}^2 - \v{T}_{a} \cdot \v{T}_{d} , \label{eq:Tad}
\end{align} 
and
\begin{align} 
    L_2(\mu_{\rm F} , k_T) = \ln\left( \frac{k_T}{\tau_1 Q} \right)^2 - \ln\left( \frac{\mu_{\rm F}}{\tau_1 Q} \right)^2.
\end{align}
If $\v{T}_{ad}^2$ were to commute with $\v{T}_s^2$, then every $i \pi$ term in Eq.~\eqref{eq:master} would cancel and there would be no coherence violation. The second equality of Eq.~\eqref{eq:Tad} is achieved by colour conservation and from it we can identify that $[\v{T}_{ad}^2,\v{T}_s^2] \neq 0$ only because of the presence of $\v{T}_{a} \cdot \v{T}_{d}$. Physically, this means that the colour charge of incoming parton $a$ can be resolved from the overall colour charge of the outgoing collinear partons it radiates due to the presence of the Coulomb exchanges. Without the $i\pi$ terms, one could identify the total charge $\v{t}_{a}^2 = \v{T}_{a}^2 +\v{T}_{d}^2 + 2 \, \v{T}_{a} \cdot \v{T}_{d}$ as the Casimir of parton ($a$), fixing $\v{t}_{a}^2 = C_{a} \v{1}_3$ and recovering the angular-ordered result \cite{Marchesini:1983bm,Webber:1983if,Marchesini:1987cf,Forshaw:2019ver,Forshaw:2021mtj}. In this sense, we can understand coherence violation as an explicit breaking of initial-state angular ordering. Whilst not phrased in the language of angular ordering, the effective field theory description of one-jettiness \cite{Stewart:2010tn}, where soft radiation is described by a low-energy matrix element of just 3 Wilson lines, similarly rests upon the identification of $\v{t}_{a}^2= C_a \v{1}_3$ so that the 4 parton colour space can be reduced to a 3 parton colour space.

With the path ordering removed, $\v{V}^{\dagger}_{\mu_{\TT{F}},k_T}\v{V}_{\mu_{\TT{F}},k_T}$ can be computed using the standard Baker-Campbell-Hausdorff expansion:
\begin{align}
    &\v{V}_{\mu_{\TT{F}},k_T}^{\dagger}\v{V}_{\mu_{\TT{F}},k_T} = \exp\Bigg( \frac{2 \As}{\pi} \v{T}_{ad}^2 L_2(\mu_{\rm F} , k_T) \\
    &  - \left(\frac{\As}{\pi}\right)^2 L_2(\mu_{\rm F} , k_T) (-i\pi) \ln\frac{k_T}{\mu_{\rm F}} [\v{T}_{ad}^2 , \v{T}_{s}^2] + \nonumber \\
    &\left(\frac{\As}{\pi}\right)^3 L_2(\mu_{\rm F} , k_T) \left( (-i\pi) \ln\frac{k_T}{\mu_{\rm F}} \right)^2 \frac{1}{3}[\v{T}_{s}^2, [ \v{T}_{s}^2 ,\v{T}_{ad}^2 ] + \dots \Bigg). \nonumber
\end{align}
To find the lowest order CVL, we expand $\v{V}^{\dagger}_{\mu_{\TT{F}},k_T}\v{V}_{\mu_{\TT{F}},k_T}$ to the first order at which the $\v{T}^2_s$ commutators do not vanish within the trace against the hard density matrix. This first happens at $\mathcal{O}(\As^3)$ giving a CVL at overall $\mathcal{O}(\As^4)$:
\begin{align}
\frac{\td \sigma_1^{\rm CVL}}{\td x_a \td x_b\td \mathcal{B}} \approx &\sum_{a = q,g}  A_{a} \left(\frac{\As}{\pi}\right)^4 \int^{Q}_{\tau_1 Q} \frac{\td k_T}{k_T} \int^{1}_{k_T/Q} ~ \frac{\td \theta}{\theta} ~ 2u(k)  ~  \nonumber \\
& \times L_2(\tau_1 Q, k_T) \left( (-i\pi) \ln\frac{k_T}{\tau_1 Q} \right)^2 , \label{eq:thelog}
\end{align}
where we have changed variables from $z$ to $\theta$ so that the boundaries of integration, enforced by $u(k)$, can be read from Fig.~\ref{fig:LundPlane}. $A_a$ contains the PDFs, hard process, and colour factors,
\begin{align}
    A_a = & f^a_A(x_a,\tau_1 Q) f^b_B(x_b,\tau_1 Q) ~ \nonumber \\
    & \times \Tr \left[\left(\frac{1}{3} [\v{T}_{s}^2, [ \v{T}_{s}^2 ,\v{T}_{ad}^2 ] \right) \v{t}_{a} \v{H}_{ab}(\mathcal{B})\v{t}^{\dagger}_{a}\right].  \label{eq:A}
\end{align}
This colour trace is non-vanishing and readily computable. Note that this is exactly the same commutator as that which appears in Eq.~(4.17) of \cite{SuperleadingLogs}. If parton $a$ is the quark and $b$ is the gluon, then
\begin{align}
    A_q&= f^q_A f^g_B \Tr \left[\left(\frac{1}{3} [\v{T}_{s}^2, [ \v{T}_{s}^2 ,\v{T}_{ad}^2 ] \right) \v{t}_{a} \v{H}_{ab}(\mathcal{B})\v{t}^{\dagger}_{a} \right]_{a={\rm quark}} \nonumber \\
    &= \frac{8}{3} \Nc^2 T_R^4 ~ f^q_A f^g_B ~ \sigma^{(0)}_{qg}(\mathcal{B}),
\end{align}
where $\sigma^{(0)}_{ab}$ is the born cross-section. If instead $a$ is the gluon:
\begin{align}
    A_g&= f^g_A f^q_B \Tr \left[\left(\frac{1}{3} [\v{T}_{s}^2, [ \v{T}_{s}^2 ,\v{T}_{ad}^2 ] \right) \v{t}_{a} \v{H}_{ab}(\mathcal{B})\v{t}^{\dagger}_{a} \right]_{a={\rm gluon}} \nonumber \\
    &= - \frac{32}{3} \Nc^2 T_R^4 ~ f^g_A f^q_B ~ \sigma^{(0)}_{gq}(\mathcal{B}).
\end{align}
We find that the CVLs are two powers of $\Nc$ suppressed, just as in the gaps-between-jets observable.

Upon performing the remaining integrals, we reveal the complete CVL:
\begin{align}
\frac{\td \sigma_1^{\rm CVL}}{\td x_a \td x_b\td \mathcal{B}} \approx & \sum_{a = q,g} A_a \left(\frac{\As}{\pi}\right)^4 (-i \pi)^2\frac{1}{480} \left(\ln\frac{1}{\tau_1}\right)^6. \label{eq:thelog}
\end{align}
Extraordinarily, we find an $\As^4 L^6$ coherence violating logarithm, which is super-leading relative to the previously known super-leading logarithms identified in gaps-between-jets.  It is simple to show that coherence violation appears at the next order as $\As^5 L^{8}$ by considering the form of $\td \sigma_2^{\rm CVL}$, where there are two unfactorised collinear emissions, and also the next term in the expansion of $\v{V}^{\dagger}_{\mu_{\TT{F}},k_T}\v{V}_{\mu_{\TT{F}},k_T}$. Again, this is super-leading relative to the previously known super-leading logarithms. We expect this behaviour to extended to higher orders as $\As^n L^{2n-2}$ for $n\geq 4$.

\section{Discussion}
The standard argument given for the presence of an $\As^4 L^5$ logarithm in the prototypical gaps-between-jets observable is that two logarithms arise from Coulomb exchanges, two logarithms come from a Sudakov double logarithm generated by a collinear emission from the initial state, and a single logarithm comes from a wide-angle-soft gluon. It is often stated that the soft gluon must be wide-angle (and therefore single logarithmic) so that the overall colour of the amplitude density matrix is off-diagonal, ensuring that the colour traces do not vanish. After all, the collinear poles in the soft-gluon antennas are colour diagonal when emitted from configurations of well-separated partons. Indeed, this fact, combined with the simplification of the soft kinematics around the collinear limit, is the backbone of QCD angular ordering. So how then do we obtain the extra logarithm?

The key point is that the additional double logarithm is not associated with collinear poles. This can be most simply appreciated by noting that we could have used a form of the antenna function in which the collinear poles have been subtracted and we would still obtain the same final result. For example, we might consider the subtraction
\begin{align}
    \omega^{\rm (sub.~ 1)}_{ij} = q_T^2 \frac{ p_i \cdot p_j}{p_i \cdot q \; p_j \cdot q} - \frac{ q_T^2 }{p_i \cdot q } \frac{T \cdot p_i}{T \cdot q} - \frac{ q_T^2 }{p_j \cdot q} \frac{T \cdot p_j}{T \cdot q} \, ,
\end{align}
where $T$ is a freely chosen (though not light-like) reference vector. Choosing $T=(1,0,0,0)$ in the lab frame, this new antenna reproduces the subtracted soft anomalous dimension ($\overline{\v{V}}$) used in the SCET approach by Becher et al. \cite{Becher:2021zkk}. An alternative, related by colour conservation when acting on a colour singlet, is
\begin{align}
    &\v{T}_i \cdot \v{T}_j \omega_{ij} \rightarrow \boldsymbol{\omega}^{\rm (sub.~ 2)}_{ij} \\
    &= q_T^2 \frac{ p_i \cdot p_j}{p_i \cdot q \; p_j \cdot q} \v{T}_i \cdot \v{T}_j \!+ \!\frac{ q_T^2 }{p_i \cdot q } \frac{T \cdot p_i}{T \cdot q} \v{T}_i^2 \!+\! \frac{ q_T^2  }{p_j \cdot q} \frac{T \cdot p_j}{T \cdot q} \v{T}_j^2 \, . \nonumber
\end{align}
Both of these subtractions cancel from the Baker-Campbell-Hausdorff commutators. This is particularly obvious in the case of $\boldsymbol{\omega}^{\rm (sub.~ 2)}_{ij}$, where it is clear that the quadratic Casimir subtractions will cancel within a colour commutator. The $\omega^{\rm (sub.~ 1)}_{ij}$ subtractions similarly cancel because they are independent of the geometry of the parent dipole. This ensures the $ad$ dipole does not vanish and colour conservation can be applied, again causing the commutators to vanish. 

Insight can also be gained by considering the Lund plane, illustrated in Fig.~\ref{fig:LundPlane}. The soft gluon (represented in red) approaches the collinear limit along the yellow boundary, and is integrated over the grey region above the dotted line and below the blue point corresponding to the collinear gluon. This grey region is an area on the plane, not a line, and consequently it integrates to give a double logarithm. However, the region exactly vanishes as it approaches the yellow line, only ever touching the line at a zero dimensional point at the tip of the triangle. Therefore, this soft gluon is being integrated over a double-logarithmically large phase-space volume which never intersects the collinear limit. Consequently, this double log is not associated with the cancellation of a $1/\epsilon^2$ pole. Rather, it is a soft-wide-angle energy logarithm with an angular coefficient that is logarithmically large. In gaps-between-jets this extra logarithm appears as a factor $Y$, the size of the rapidity gap.

We conclude our discussion by noting the role played by the choice of ordering variable. It is known that the calculation of super-leading logarithms in the gaps-between-jets observable, using a parton branching approach, is sensitive to the choice of ordering variable, and that it is necessary to employ the dipole $k_{T}$ ordering~\cite{SuperleadingLogs}, or an equivalent $k_{T}$-type ordering that simultaneously regularises both soft and collinear divergences~\cite{ColoumbGluonsOrdering}. We note that for this observable, the leading $\As^4 L^{6}$ CVL is recovered using lab-frame $k_T$ though this agreement does not persist to order $\As^4 L^{5}$.

The presence of an unexpected additional logarithm changes the overall log counting for coherence violation in global observables (first given in \cite{Banfi:2010xy} and expanded upon in \cite{Forshaw:2021fxs}). Specifically, we have shown that coherence violation may be further enhanced by one more logarithm than naive expectations depending on the angular boundary defined by the observable. For $n$-jettiness, and other forward suppressed observables, coherence violation is a leading-log effect that appears at next-to-next-to double log in the expansion ($\As^n L^{2n-2}$ for $n\geq 4$).

The fact that one-jettiness suffers an $\As^4 L^6$ coherence violating logarithm raises several questions for future studies. Firstly, 
recent years have seen a great deal of success resumming the super-leading logarithms in gaps-between-jets \cite{Becher:2021zkk,Becher:2023mtx,Boer:2023jsy,Boer:2023ljq,Boer:2024hzh,Boer:2024xzy,Becher:2024kmk,Becher:2025igg,Banerjee:2025kkq} and their effect can be sizeable. However, key to this success has been that the soft-gluon is single logarithmic and so only a single soft insertion is required. This helps to ensure that the colour structures are analytically tractable. Jettiness violates this paradigm to an extreme degree. It will therefore require new analytical developments to handle the large multiplicities of soft gluons needed to perform the resummation. Alternatively, numerical codes that can account for a large number of soft gluons, might be able to handle this resummation \cite{Forshaw:2025bmo,Forshaw:2025fif}. It also remains unanswered how factorisation breaking impacts soft-sensitive, non-IRC-safe observables, which are sometimes used to study fragmentation and TMD physics \cite{Boussarie:2023izj}. In particular, we have shown that the $\As^4 L^6$ term in jettiness is not associated with an additional soft-collinear pole but rather it reflects the presence of a logarithmically enhanced phase-space volume. This is consistent with the result found in \cite{Becher:2024kmk,Becher:2025igg} where it was demonstrated that, within an EFT perspective, there are no extra poles in low energy matrix elements below the soft scale for gaps-between-jets, ensuring parton distribution functions (PDFs) remain well defined \cite{Collins:1987pm,Collins:1988ig}.

\section{Acknowledgements}
The authors would like to thank the Mainz Institute of Theoretical Physics (MITP) for hospitality and support. MITP has received funding from the Cluster of Excellence PRISMA+ (EXC 2118/1, Project ID 390831469) funded by the German Research Foundation (DFG). AB acknowledges the hospitality of the Department of Physics of Royal Holloway University of London, where part of this work was performed. JH is supported by the Leverhulme Trust as an Early Career Fellow. AB is supported by the UK Science and Technology Facility Council under the grant ST/X000796/1 and JF is supported by the UK Science and Technology Facility Council under the grant ST/X00077X/1.

\bibliography{jettiness}

\end{document}